\documentclass[10pt,prd,preprintnumbers,floatfix,aps,nofootinbib,showpacs,twocolumn,superscriptaddress,noshowpacs]{revtex4-1} 
\usepackage[latin9]{inputenc}
\usepackage[dvipsnames]{xcolor}
\usepackage{graphicx}
\usepackage{hyperref}
\usepackage{amsthm,amsmath,amssymb,amsfonts}
\usepackage{amsbsy}
\usepackage{bbm}
\usepackage{mathtools}
\usepackage{microtype}

\newcommand{\SK}[1]{\textcolor{purple}{{#1}}}

\makeatletter

\pdfpageheight\paperheight
\pdfpagewidth\paperwidth

\@ifundefined{textcolor}{}{%
 \definecolor{BLACK}{gray}{0}
 \definecolor{WHITE}{gray}{1}
 \definecolor{RED}{rgb}{1,0,0}
 \definecolor{GREEN}{rgb}{0,1,0}
 \definecolor{BLUE}{rgb}{0,0,1}
 \definecolor{CYAN}{cmyk}{1,0,0,0}
 \definecolor{MAGENTA}{cmyk}{0,1,0,0}
 \definecolor{YELLOW}{cmyk}{0,0,1,0}
}

\makeatother

\makeatletter

\DeclareRobustCommand{\rcite}[1]{%
  \rcite@aux#1,\@nil{#1}%
}
\def\rcite@aux#1,#2\@nil#3{%
  \if\relax#2\relax
    Ref.~\cite{#3}%
  \else
    Refs.~\cite{#3}%
  \fi
}
\makeatother
\definecolor{wine}{RGB}{136,34,85}
\definecolor{teal}{RGB}{0,85,102}
\hypersetup{
    colorlinks = true,
    citecolor = {wine},
    linkcolor = {teal},
    urlcolor = {teal},
}

\newcommand{\be}{\begin{equation}}
\newcommand{\ee}{\end{equation}}
\newcommand{\ba}{\begin{eqnarray}}
\newcommand{\ea}{\end{eqnarray}}

\newcommand{\dd}{\mathrm{d}}

\newcommand{\lcdm}{$\Lambda$CDM}

\begin{document}

\preprint{IFT-UAM/CSIC-24-119}

\title{DESI constraints on $\alpha$-attractor inflationary models}

\author{George Alestas}
\email{g.alestas@csic.es}
\affiliation{Instituto de F\'isica Te\'orica (IFT) UAM-CSIC, C/ Nicol\'as Cabrera 13-15, Campus de Cantoblanco UAM, 28049 Madrid, Spain}

\author{Marienza Caldarola}
\email{marienza.caldarola@csic.es}
\affiliation{Instituto de F\'isica Te\'orica (IFT) UAM-CSIC, C/ Nicol\'as Cabrera 13-15, Campus de Cantoblanco UAM, 28049 Madrid, Spain}

\author{Sachiko Kuroyanagi}
\email{sachiko.kuroyanagi@csic.es}
\affiliation{Instituto de F\'isica Te\'orica (IFT) UAM-CSIC, C/ Nicol\'as Cabrera 13-15, Campus de Cantoblanco UAM, 28049 Madrid, Spain}
\affiliation{Department of Physics and Astrophysics, Nagoya University, Nagoya, 464-8602, Japan}

\author{Savvas Nesseris}
\email{savvas.nesseris@csic.es}
\affiliation{Instituto de F\'isica Te\'orica (IFT) UAM-CSIC, C/ Nicol\'as Cabrera 13-15, Campus de Cantoblanco UAM, 28049 Madrid, Spain}

\date{\today}

\begin{abstract}
The recent results on the baryon acoustic oscillations measurements from the DESI collaboration have shown tantalizing hints for a time-evolving dark energy equation of state parameter $w(z)$, with a statistically significant deviation from the cosmological constant and cold dark matter (\lcdm) model. One of the simplest and theoretically well-motivated plausible candidates to explain the observed behavior of $w(z)$, is scalar-field quintessence. Here, we consider a class of models known as $\alpha$-attractor, which describe in a single framework both inflation and the late-time acceleration of the Universe. Using the recent DESI data, in conjunction with other cosmological observations, we place stringent constraints on $\alpha$-attractor models and compare them to the \lcdm\, model. We find the $\alpha$ parameter of the theory, which is physically motivated from supergravity and supersymmetry theories to have the values $3\alpha \in \{1,2,3,4,5,6,7\}$, is constrained to be $\alpha\simeq 1.89_{-0.35}^{+0.40}$. In addition, we find that the rest of the cosmological parameters of the model agree with the corresponding values of \lcdm, while a Bayesian analysis finds strong support in favor of the $\alpha$-attractor model. We also highlight an interesting connection between the $\alpha$-attractor models and the stochastic gravitational wave background, where a contribution to the latter could derive from an enhancement of inflationary gravitational waves at high frequencies due to an early kination phase, thus providing an interesting alternative way to constrain the theory.
\end{abstract}

\maketitle
%

\section{Introduction \label{sec:introduction}}
One of the most intriguing questions in theoretical physics for the past thirty years or so, is the observed accelerated expansion of the Universe at late times. The existence of this phenomenon has been confirmed using a plethora of different cosmological probes, including Type Ia Supernovae, the cosmic microwave background (CMB) anisotropies, and large scale structure (LSS) probes such as the baryon acoustic oscillations (BAO). This accelerated expansion implies the existence of a repulsive force that dominates over gravity on cosmological (large) scales and within the framework of general relativity (GR), it implies the existence of fluid with a negative equation of state, commonly dubbed as dark energy (DE). Currently, the consensus is that this era of accelerated expansion of the Universe is due to the presence of a cosmological constant $\Lambda$, which behaves as a uniform vacuum energy. This model is in excellent agreement with the observations~\cite{Planck:2018vyg}. 

Still, despite the success of this simple model, several tensions have recently appeared between low redshift and high redshift probes, see for example Ref.~\cite{Perivolaropoulos:2021jda}. Moreover, the cosmological constant also suggests fine-tuning as its value differs by orders of magnitude with what is predicted by quantum field theories~\cite{Weinberg:1988cp,Carroll:2000fy}, faces various theoretical problems \cite{Martin:2012bt}, and is also plagued by the so-called coincidence problem \cite{Zlatev:1998tr}. Specifically, Ref.~\cite{Martin:2012bt} discusses several aspects of the cosmological constant model, not only in the original version discussed in Ref.~\cite{Weinberg:1988cp}, and that in reality the renormalized value of the zero-point energy density is in fact far from being 122 orders of magnitude larger than the critical energy density, as often mentioned in the literature. These issues stimulated the creation of several DE models based on ad-hoc modification of gravity or scalar fields that mediate the force between particles, e.g. canonical scalar fields~\cite{Ratra:1987rm,Wetterich:1987fm,Caldwell:1997ii}, scalar fields with generalized kinetic terms~\cite{ArmendarizPicon:2000dh,ArmendarizPicon:2000ah, Chiba:1999ka}, non-minimal couplings~\cite{Uzan:1999ch,Perrotta:1999am,Riazuelo:2001mg, Chiba:1999wt} or coupled DE models~\cite{Dent:2008vd}, in addition to GR.

Out of all the aforementioned candidates, arguably one of the simplest and most well-known candidates for DE is quintessence~\cite{Tsujikawa:2013fta}, namely a single, canonical, light and slowly rolling scalar field, leading to an accelerated expansion~\cite{Amendola:2015ksp}. 
By tuning the properties of the scalar field, e.g. the potential or its kinetic term, the scalar field can control the fate of both the early and late Universe by dominating its energy density at either time.

As mentioned, scalar fields can also be used in early time physics, to describe inflation, a period of exponential accelerated expansion of the Universe right after the Big Bang. Inflation was originally introduced to provide an explanation to several problems in cosmology~\cite{GuthPhysRevD.23.347}, such as the horizon problem, related to the fact that CMB radiation has almost exactly the same temperature across the sky. Inflation also provides a solution for generating primordial density perturbations, the seeds of structure formation, by amplifying initial quantum fluctuations. This variation in matter density eventually led to denser regions clustering, leading to galaxy formation~\cite{Mukhanov:1981xt}.

In the most general picture of single-field slow-roll inflation, the inflaton evolves along a flat region of its potential, slowly rolling and driving cosmic acceleration. As the inflaton exits this flat region, it begins to move faster, leading to the end of inflation. While single-field slow-roll inflation successfully addresses classical cosmological problems, several intriguing features emerge when moving beyond this simplified model. These include distinguishing effective field theories within the string landscape~\cite{Achucarro:2018vey}, producing primordial black holes without having a strong scale dependence at small scales~\cite{Palma:2020ejf,Fumagalli:2020adf}, and possible violations of the consistency relations for local primordial non-Gaussianity~\cite{Cabass:2022ymb}. 

One class of models that extends beyond the simple inflationary scenario is the so-called $\alpha$-attractor models, which link early- and late-time physics~\cite{Kallosh:2013yoa, Galante:2014ifa, Linder:2015qxa, Geng:2015fla,Ueno:2016dim,Shahalam:2016juu, Braglia:2020bym,DIMOPOULOS2002287,Dimopoulos:2017zvq,Garc_a_Garc_a_2018,LinaresCedeno:2019bgo,de_Haro_2021,Akrami_2021,Akrami:2017cir,AresteSalo:2021wgb,Bhattacharya:2022akq, Rodrigues:2020fle}. In this scenario, the inflaton is a scalar field driving inflation, whose main properties, such as duration and spectrum of primordial perturbations, are characterized by the inflaton field itself and the shape of its potential. In the context of $\alpha$-attractor models, the $\alpha$ parameter determines the shape of the potential. These models of inflation are compelling since they are in accordance with current CMB observations and can be also embedded in supergravity and string theories~\cite{Ferrara:2016fwe,Kallosh:2017wku,Scalisi:2018eaz,AresteSalo:2021wgb}.

Cosmological observations are particularly well-suited to constrain these models, as both early- and late-time physics can leave imprints in the LSS and the CMB (both measured by dedicated surveys), thus allowing us to probe for hints of these models. In this regard, the Dark Energy Spectroscopic Instrument (DESI) is particularly useful for probing the physics of DE at late times, by mapping the LSS of the Universe. Recently, the collaboration released the first year data~\cite{DESI:2024uvr, DESI:2024mwx, DESI:2024lzq}, where they report the latest results from BAO measurements and the resulting cosmological constraints. Interestingly, a statistically significant deviation from \lcdm\ was also reported, which could potentially be explained by a quintessence model~\cite{Tada:2024znt,Yin:2024hba,Bhattacharya:2024hep,Ramadan:2024kmn,Wolf:2024eph,Giare:2024gpk} or systematics in the DESY3 data \cite{RoyChoudhury:2024wri}. 

In this paper, we use the latest observational CMB and LSS data, which provide information on early- and late-time physics respectively, to place constraints on the $\alpha$-attractor models and compare them to the \lcdm, model. This allows us to test the compatibility of these models with the latest cosmological data and provide a unified explanation for both inflation and DE. In this context, we also explore the connections between the $\alpha$-attractor model and gravitational waves (GWs)~\cite{WaliHossain:2014usl,Dimopoulos:2017zvq,Ahmad:2019jbm,Wang:2024sgo}. A key feature of quintessential inflation is the post-inflation phase dominated by the scalar field's kinetic energy. This alters the Hubble evolution, enhancing inflationary GWs at high frequencies compared to the standard radiation-dominated Universe~\cite{Giovannini:1998bp,Peebles:1998qn,Giovannini:1999bh,Giovannini:1999qj,Giovannini:2008tm,Tashiro:2003qp,Figueroa:2018twl,Figueroa:2019paj,Duval:2024jsg}. To investigate this, we first estimate the amplitude of the GW spectrum based on the parameter values inferred from CMB and LSS data. We then discuss constraints on these parameters using the current upper bounds on the GW amplitude from big-bang nucleosynthesis (BBN) and the CMB, as well as theoretical limits related to the allowed duration of the kination phase, which is tied to the reheating temperature.

Our paper is organized as follows: in Sec.~\ref{sec:alpha_model}, we provide an overview for the $\alpha$-attractor models of inflation. Then, in Sec.~\ref{sec:data}, we provide a description of cosmological data used to place constraints on the model, while in Sec.~\ref{sec:results}, we present and discuss our results. In Sec.~\ref{sec:GWs}, we analyse the connection of $\alpha$-attractor models with GWs, deriving the predicted spectrum of inflationary GWs in such models and evaluating the allowed region that fall within the sensitivity of current and future GW experiments. Finally, we conclude in Sec.~\ref{sec:conclusions}.

\section{$\alpha$-attractor models \label{sec:alpha_model}}
Here, we provide a brief overview of the theory of the $\alpha$-attractor models. Specifically, we consider an action of the form
\begin{equation}
\label{eq:action}
S = \int \dd^4x\, \sqrt{-g}\,\left[\frac{1}{2}M_{\rm Pl}^2R-\frac{1}{2}\frac{\partial_\mu\phi\,\partial^\mu\phi}{\left(1-\frac{\phi^2}{6\alpha}\right)^2}-V(\phi) + \mathcal{L}_\text{m}\right],
\end{equation}
where we assume natural units with $\hbar=c=1$, $M_{\rm Pl}=1/\sqrt{8\pi G}$ is the reduced Planck mass, $R$ is the usual four-dimensional Ricci scalar, $V(\phi)$ is the potential of the scalar field $\phi(\vec{x},t)$, and $\mathcal{L}_\text{m}$ is the matter Lagrangian density including the Standard Model particles. 
Note that we rescale the field $\phi$ by $M_{\rm Pl}$ and the curvature parameter $\alpha$ by $M_{\rm Pl}^2$.
The main parameter in this configuration is $\alpha$, which is related to the curvature of the field space and is typically assumed to be $\alpha\sim\mathcal{O}(1)$.\footnote{This $\alpha$ is not to be confused with the fine-structure constant $\alpha_\mathrm{EM}\sim1/137$.} 

A key feature of $\alpha$-attractor models is that the kinetic term in the original Lagrangian has a pole at $\phi = \pm \sqrt{6\alpha}$. One can redefine the scalar field in terms of a canonically normalized field $\varphi$, as
\begin{equation}
\label{eq:canonical_transformation_phi}
    \phi = \sqrt{6\alpha} \tanh\frac{\varphi}{\sqrt{6\alpha}}.
\end{equation}
With this transformation, this pole is shifted to infinity at the cost of introducing an inflection point in the potential. When $\alpha\rightarrow \infty$ then the two fields, namely $\phi$ and $\varphi$ become equal, while in general in terms of $\varphi$, the potential becomes sufficiently flat to support accelerated expansion. It is possible to consider different functions to build up the inflaton potential, but it is interesting to focus on functions that allow for the presence of an inflection point.

Based on the discussion in Ref.~\cite{Akrami_2021,Akrami:2017cir,Dimopoulos:2017zvq}, we consider the following exponential potential, which naturally arises in high-energy theories such as supergravity or string theory:
\be
V(\phi)= M^2e^{\gamma\left(\frac{\phi}{\sqrt{6\alpha}}-1\right)} + V_0.
\ee
By applying the transformation of Eq.~\eqref{eq:canonical_transformation_phi}, we obtain
\be
V(\varphi)= M^2e^{\gamma\left(\tanh\frac{\varphi}{\sqrt{6\alpha}}-1\right)} + V_0. \label{eq:pot_a}
\ee
In our analysis, we set $V_0=0$. In the context of $\alpha$-attractor models, $\alpha$ is a free parameter that can be constrained phenomenologically. Nevertheless, interesting theoretical values of $3\,\alpha \in \{1,2,3,4,5,6,7\}$, are set by supersymmetry and supergravity, as reported in Refs.~\cite{Ferrara:2016fwe, Kallosh:2017ced, Kallosh:2017wnt}. 

In the asymptotic regime, for large and positive $\varphi$, where we assume inflation occurs, the effective potential is approximated by
\begin{equation}
V(\varphi) \simeq M^2\left(1-2\gamma e^{-\frac{2\varphi}{\sqrt{6\alpha}}}\right)+V_0 \,.
\end{equation}
With this approximated form, the number of e-folds can be calculated using the slow-roll approximation, which gives $N_\ast \simeq \int_{\varphi_*}^{\varphi_{\rm end}} ~ d\varphi V/V_{,\varphi}$ with $\varphi_\ast$ and $\varphi_{\rm end}$ being the value of the scalar field when the corresponding mode exits the Hubble radius and at the end of inflation, respectively. Keeping only the leading-order term, we obtain 
\begin{equation}
N_\ast \simeq \frac{3\alpha}{4\gamma}
e^{-{\frac{2\varphi_\ast}{\sqrt{6\alpha}}}} \,.
\label{eq:Nefold}
\end{equation}
Using this, $M$ can be related to $\mathcal{A}_s$, the amplitude of the primordial scalar power spectrum, as 
\begin{equation}
\frac{M^2}{M_{\rm Pl}^4} = \frac{144\pi^2\alpha N_{*}}{(2N_{*}-3\alpha)^3}\,\mathcal{A}_s.
\label{eq:M2}
\end{equation}
 
The scalar spectral index $n_s$ and the tensor-to-scalar ratio $r$ are related to the number of e-folds of inflation $N_{*}$ and the parameter $\alpha$ via
\begin{equation}
n_s = 1-\frac{2}{N_{*}}, \qquad r=\frac{12\alpha}{N_{*}^2}. \label{eq:ns_r}
\end{equation}
Furthermore, in this framework the late-time values of the DE equation of state parameter $w$ are also related to the primordial ones. In particular, one finds~\cite{Zhumabek:2023wka}
\ba 
w_0 &=& -1+\frac{4}{3N_{*}^2\, r}, \label{eq:w0}\\
w_a &\approx & -\frac{2}{3N_{*}^2\, r},\label{eq:wa}
\ea 
where $(w_0,w_a)$ are the values of the DE equation of state parameter $w(a)$ and its derivative today. Of particular interest though, in light of the DESI data, are high values of $\alpha\sim7/3$ as in this case we find the effective equation of state parameters to be $(w_0,w_a)\simeq(-0.95,-0.06)$.

In the following analysis, we do not use the approximated forms, Eqs.~\eqref{eq:w0}-\eqref{eq:wa}, but instead we choose to solve the full dynamical equations.
In this case, assuming the flat Friedmann-Lema\'itre-Robertson-Walker (FLRW) metric, the background dynamics at late times can be described by
\ba
&& H^2\left(3-\frac12 \varphi'^2\right) M_{\rm Pl}^2 = V(\varphi)+\rho_\mathrm{m}+\rho_\mathrm{r}, \label{eq:H2} \\
&& \varphi''+\left(3+\frac{H'}{H}\right)\varphi'+H^{-2}\,\frac{\dd V(\varphi)}{\dd \varphi} = 0,\label{eq:f2}
\ea 
where $\rho_\mathrm{m}$ and $\rho_\mathrm{r}$ are the matter (baryons $\rho_\mathrm{b}$ plus cold dark matter $\rho_\mathrm{cdm}$) and radiation energy densities, respectively, primes are derivatives with respect to the number of e-folds $N=\ln(a)$, $H=\dot{a}/a$ is the Hubble parameter, and a dot is a derivative with respect to the cosmic time $t$.

The scalar field freezes after reheating and begins to move again once it overcomes Hubble friction. To calculate the late-time dynamics of the scalar field, we set the initial condition to $\varphi_\mathrm{init}=-10$. Note that the value of the initial $\phi$ is not very important, one just needs to make sure the field starts rolling after the freeze epoch and before the DE domination phase. For comparison reasons we used the same value as in Ref.~\cite{Akrami_2021}.

For comparison, we also use the flat \lcdm\, model, in which the expansion history is simply described by the usual Friedmann equation
\be
3H^2 M_{\rm Pl}^2 = \rho_\mathrm{m}+\rho_\mathrm{r}+\rho_\mathrm{\Lambda}, \label{eq:H2L}
\ee 
where the cosmological constant energy density is given by $\rho_\mathrm{\Lambda}=3\Lambda$. 
The energy density parameters are defined by $\Omega_x\equiv \rho_x/(3 M_{\rm Pl}^2 H_0^2)$ for each component.

Following Ref.~\cite{Akrami_2021}, we focus on the effects of the $\alpha$-attractor models at late times, assuming the early and late-time physics are connected via the relations given by Eqs.~\eqref{eq:M2}-\eqref{eq:ns_r}. Then, we modify the publicly available Boltzmann code \texttt{CLASS}~\cite{Lesgourgues:2011re, Blas:2011rf}, which already includes a module for solving the quintessence equations at late times, and we include the potential of Eq.~\eqref{eq:pot_a} along with the parameters that connect early and late times via Eqs.~\eqref{eq:M2}-\eqref{eq:ns_r}. For our analysis, we set $V_0=0$ in Eq.~\eqref{eq:pot_a}. 

It should be noted that the parameter $\gamma$ that appears in Eq.~\eqref{eq:pot_a} is not free to vary, but instead is determined and automatically adjusted using shooting, such that the Friedmann equation Eq.~\eqref{eq:H2} is satisfied today, i.e., $H(a=1)=H_0$. The reason for employing the shooting method is that, in an MCMC approach, a point is selected a priori in the parameter space, and the process jumps to that point where we calculate the observables using the \texttt{CLASS} code. During this process, one has to solve both the Friedmann and Klein-Gordon equations simultaneously, and the condition $H(a=1)/H_0=1$ might not be satisfied for the given set of parameter values. Consequently, we need to transform the boundary value problem, which has conditions set at early times and requires $H(a=1)/H_0=1$, into a pure initial condition problem that inherently satisfies all the required conditions. In our case, we have to adjust one of the parameters of the model, either $\Omega_{\rm m}$ or $\alpha$ or $\gamma$. Given that $\gamma$ is the least physically motivated parameter, we select it for the shooting.

\section{Observational data and methodology\label{sec:data}}
In our analysis, we use three datasets comprising of CMB, BAO, and supernovae measurements. More specifically, we use the full Planck 2018 data, i.e., the TT,TE,EE,lowE and Lensing likelihoods~\cite{Planck:2018vyg}, hereby referred to as CMB$+$Lensing, and the newly released DESI DR1 BAO measurements~\cite{DESI:2024mwx} (for an alternative study on BAO measurements see \cite{Anselmi:2022exn, ODwyer:2019rvi, Anselmi:2018vjz}). Lastly, we also use the Pantheon$+$ Type Ia supernovae (SnIa) data~\cite{Riess:2021jrx}. 

We consider two distinct cases: the vanilla flat \lcdm\, model, which is characterized by five key cosmological parameters $\{ \omega_b, \Omega_\mathrm{m}, n_s, \mathcal{A}_s, H_0 \}$ with $\omega_\mathrm{b}\equiv\Omega_\mathrm{b} h^2$, and the $\alpha$-attractor model, which has six key parameters $\{\omega_\mathrm{b}, \Omega_\mathrm{m}, n_s, \alpha, \mathcal{A}_s, H_0 \}$. 
In the $\alpha$-attractor model, the theory is characterized by three fundamental physical parameters: $\alpha$, $M$, and $N_*$. Here, we treat $\alpha$ as a free parameter to be constrained, while $M$ and $N_*$ can be related to the two CMB parameters, $n_s$ and $\mathcal{A}_s$, via Eqs.~(\ref{eq:M2}) and (\ref{eq:ns_r}). Although parameter constraints could be expressed in terms of the fundamental physical parameters, for consistency and direct comparison with the \lcdm\ model, we adopt the latter parameter set. As a result, the $\alpha$-attractor model can be viewed as a one-parameter extension of the \lcdm\ framework, with $n_s$ and $\mathcal{A}_s$ implicitly encoding information about $M$ and $N_*$.

The uncertainties in other free parameters, such as the optical depth $\tau$ and the Planck nuisance parameters, which are non-cosmological and related to the instrument, are marginalized over in the results presented in the next section. We consider reasonable priors for all the standard \lcdm\ parameters to encompass a broad parameter space, while we choose $\alpha \in[0.01, 5]$ and $\alpha \in[0.5, 3.5]$ for the CMB$+$Lensing and the full data compilation respectively.

It should be noted that our analysis is different from that of Ref.~\cite{Giare:2024sdl}, as we include the dynamical equations of the $\alpha$-attractor models and not only the $(w_0, w_a)$ values for the DE equation of state. Furthermore, in our work we use the more recent DESI data. Nevertheless, we have checked that in the cases where our analyses overlap, they are in good agreement.

Then, for the actual runs we perform a Markov-chain Monte-Carlo (MCMC) analysis for both models, assuming the free parameters in each model, plus the various nuisance parameters of the likelihoods. Specifically, we use a Metropolis-Hastings MCMC algorithm, using the publicly available \texttt{MontePython} code~\cite{Brinckmann:2018cvx}. For the Boltzmann code, we use a modified version of the publicly available \texttt{CLASS} code~\cite{Lesgourgues:2011re,Blas:2011rf}.\footnote{Our modified \texttt{CLASS} version can be found at \href{https://github.com/snesseris/class_alpha_attractor}{https://github.com/snesseris/class\_alpha\_attractor}.} Note that our modified \texttt{CLASS} version numerically solves the full scalar field equations Eqs.~\eqref{eq:H2}-\eqref{eq:f2} and does not just use the simple $w_0$-$w_a$ parameters mentioned earlier.

\section{MCMC analysis \label{sec:results}}
Now we present the results of our MCMC analysis for the two models we presented earlier. In particular, in Tables~\ref{tab:MCMC_bf_cmb} and \ref{tab:MCMC_bf_full}, we show the 68.3\% confidence limits of the cosmological parameters derived by the MCMCs for the two cases. As seen in the tables, all the parameters shared between the two models show good agreement. On the other hand, for the $\alpha$ parameter, we obtain an upper limit of $\alpha<3$ with only the CMB data, while we find the constraint $\alpha\simeq 1.89_{-0.35}^{+0.40}$, also including the DESI data.

\begin{figure*}[t!]
\includegraphics[width=1\textwidth]{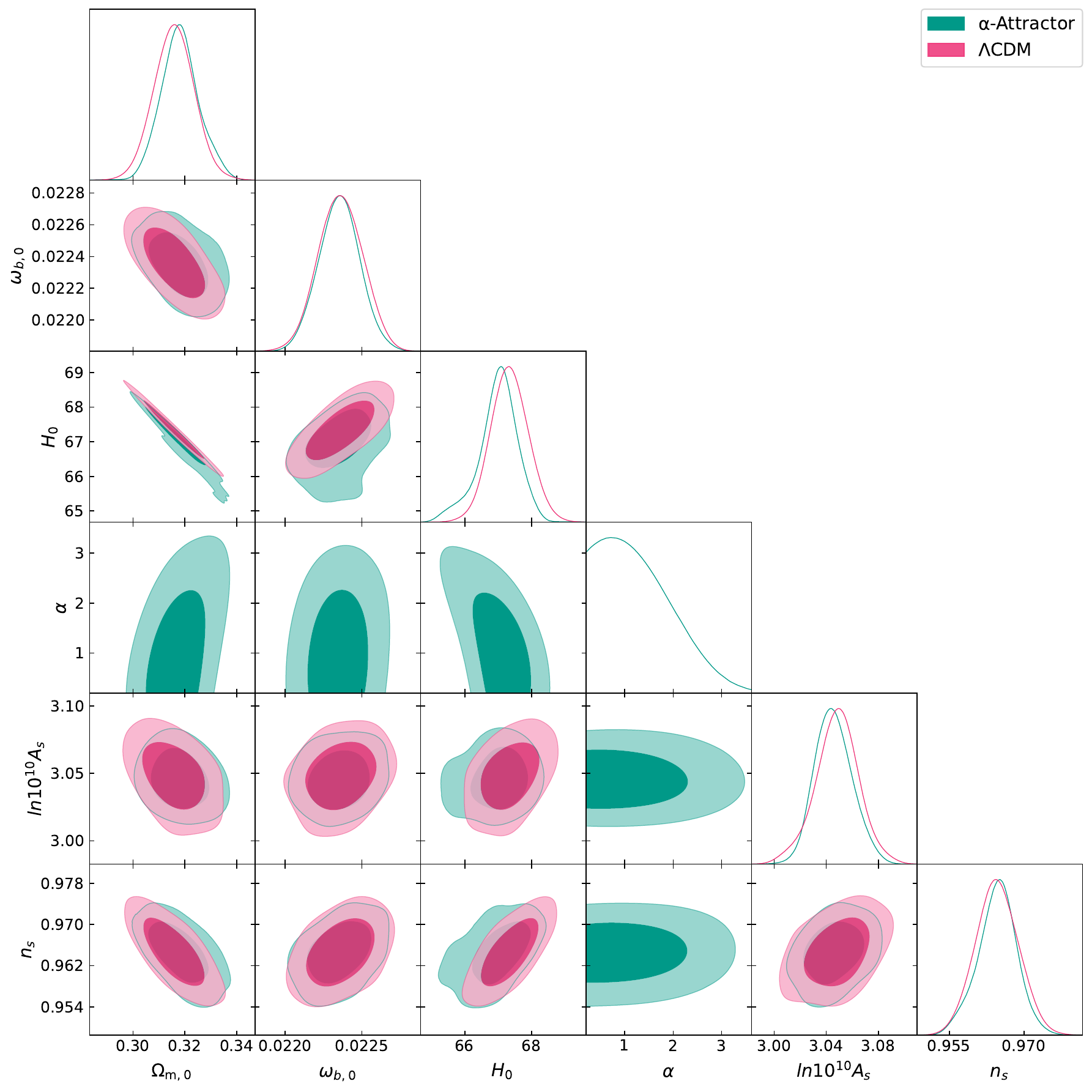}
\caption{\label{fig:cmb}The $68.3 \%$--$95.5 \%$ confidence contours for the cosmological parameters of the \lcdm\, and $\alpha$-attractor models, which  are derived using CMB$+$Lensing data.}
\end{figure*}

\begin{figure*}[t!]
\includegraphics[width=1\textwidth]{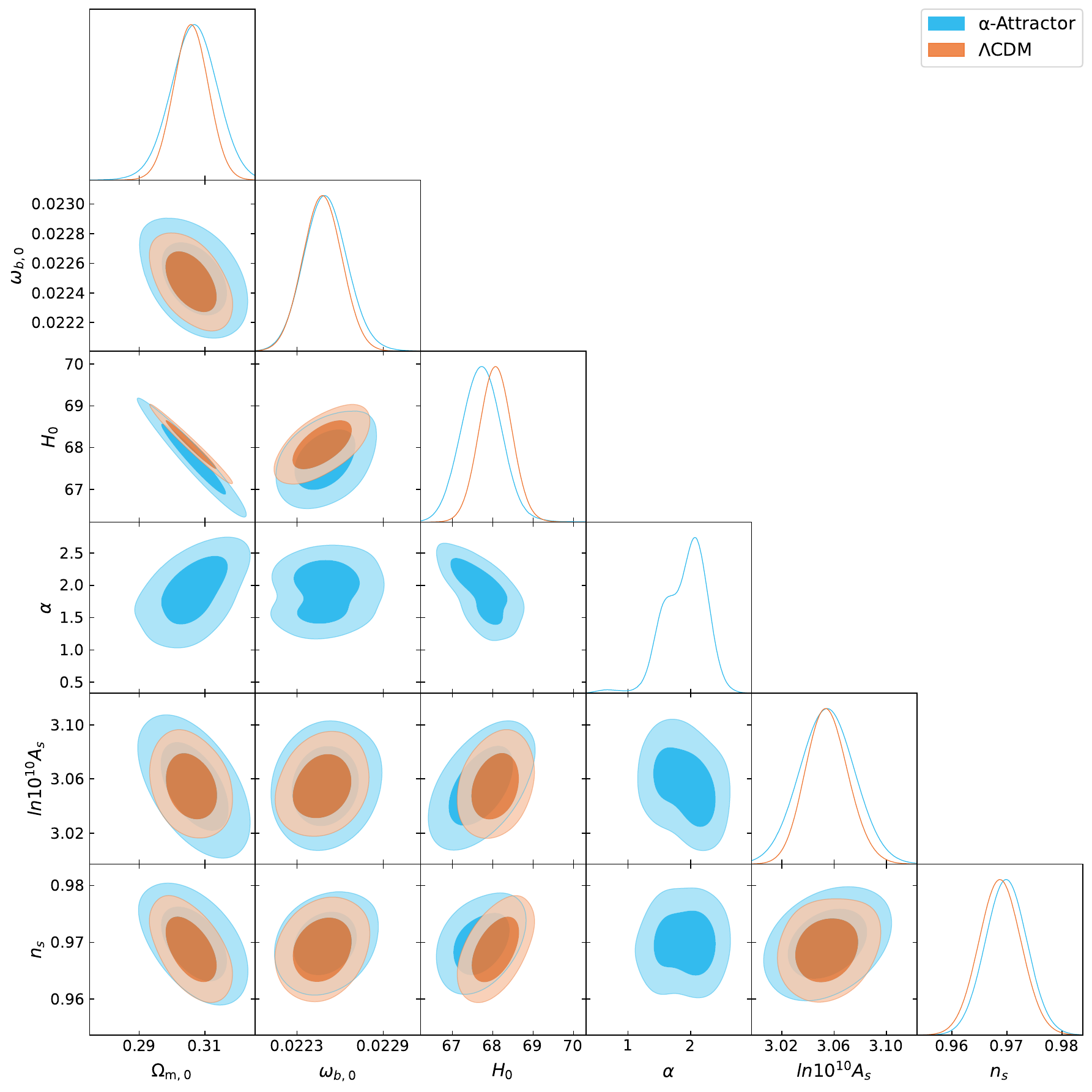}
\caption{\label{fig:full}The same as in Fig. \ref{fig:cmb}, but this time a combination of CMB$+$Lensing, DESI and Pantheon$+$ likelihoods are used.}
\end{figure*}

In Figs.~\ref{fig:cmb} and \ref{fig:full}, we show the $68.3 \%$--$95.5 \%$ confidence contours for a subset of the cosmological parameters of the $\alpha$-attractor model. More specifically, Fig.~\ref{fig:cmb} corresponds to CMB$+$Lensing data, while Fig.~\ref{fig:full} corresponds to the CMB$+$Lensing, DESI and Pantheon$+$ likelihoods. Here we find a constraint on $\alpha$ when all data are used. On the contrary, the parameter was found to be unconstrained in Ref.~\cite{Giare:2024sdl}, using the simpler analysis with the $w_0$-$w_a$ parameters. Again,  for the remaining parameters, we find good agreement with those of the \lcdm\,  model. 

The best-fit value for $\alpha$, obtained when all datasets are used, corresponds to an evolving DE equation of state parameter, $w(z)$, and is generally consistent with Ref.\cite{DESI:2024mwx}. Specifically, $\alpha$ is related to  $(w_0,w_a)$ via Eqs.~\eqref{eq:w0}-\eqref{eq:wa}, thus having an effect at late times in the ISW effect (see Appendix~\ref{sec:app}). When DESI is added, which mainly constrains the angular diameter distance via the BAO measurements, again the effect is around $\sim6\%$ at $z\sim1$ for $\alpha=4$, see Fig.~\ref{fig:DA} in Appendix~\ref{sec:app}. 

\begin{figure}[t!]
\includegraphics[width=1\columnwidth]{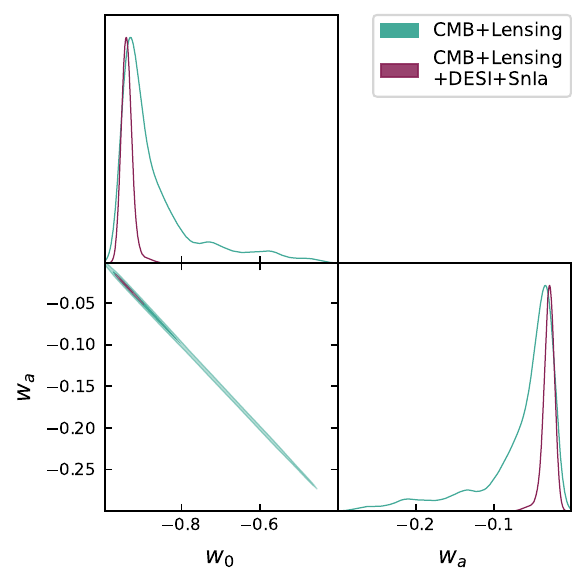}
\caption{\label{fig:attractorw0wa}The $68.3 \%$--$95.5 \%$ confidence contours for the $w_0$ and $w_a$ parameters of the $\alpha$-attractor model as defined in Eqs.~\eqref{eq:w0} and \eqref{eq:wa}. The green contours correspond to CMB$+$Lensing data while the purple contours use CMB$+$Lensing, DESI and Pantheon$+$ likelihoods. As seen, the $w_0-w_a$ contour is in good agreement (in terms of the orientation and shape) with Fig. 6 (right panel) of Ref.~\cite{DESI:2024mwx}, albeit our contour is of course smaller as in the $\alpha$-attractor model the two parameters are related via Eqs.~\eqref{eq:w0}-\eqref{eq:wa}.}
\end{figure}

To further expand on this, in Fig.~\ref{fig:attractorw0wa} we show the $68.3 \%$--$95.5 \%$ confidence contours for the $w_0$ and $w_a$ parameters of the $\alpha$-attractor model, defined in Eqs. \eqref{eq:w0} and \eqref{eq:wa}. The cyan contours correspond to CMB$+$Lensing data while the orange ones use the combination CMB$+$Lensing, DESI and Pantheon$+$ likelihoods. As can be seen, the contours follow the roughly linear relationship between the parameters, see also via Eqs.~\eqref{eq:w0}-\eqref{eq:wa}, while remaining in broad agreement (in terms of orientation and shape) with the DESI $(w_0,w_a)$ contours with the same data combination (see the right panel in Fig. 6 of  Ref.~\cite{DESI:2024mwx}).

\begin{table}[t!]
\centering
\begin{tabular}{|c|c|c|}
\hline
Parameters & \lcdm & $\alpha$-attractor \\
\hline
$\Omega_{\rm m}$ & $0.317^{+0.015}_{-0.008}$ & $0.319_{-0.008}^{+0.007}$  \\
$\omega_b$ & $0.0224_{-0.0001}^{+0.0002}$ & $0.02224^{+0.0003}_{-0.0002}$  \\
$H_0$ & $67.3^{+1.1}_{-0.6}$ & $66.9_{-0.6}^{+0.7}$  \\
$\alpha$ & N/A & $<3$  \\
$\ln 10^{10}A_s$ & $3.047^{+0.0003}_{-0.0002}$ & $3.048_{-0.015}^{+0.013}$  \\
$n_s$ & $0.964^{+0.009}_{-0.004}$ & $0.965_{-0.004}^{+0.004}$  \\
\hline
\end{tabular}
\caption{Constraints at 68.3\% confidence limits of the cosmological parameters for the $\alpha$-attractor and \lcdm\, models, derived using CMB and Lensing data.}
\label{tab:MCMC_bf_cmb}
\end{table}

\begin{table}[t!]
\centering
\begin{tabular}{|c|c|c|}
\hline
Parameters & \lcdm & $\alpha$-attractor \\
\hline
$\Omega_{\rm m}$ & $0.309_{-0.008}^{+0.003}$ & $0.305_{-0.004}^{+0.008}$  \\
$\omega_b$ & $0.0224_{-0.0001}^{+0.0002}$ &  $0.0225\pm 0.0001$ \\
$H_0$ & $67.9_{-0.2}^{+0.6}$ & $67.9_{-0.7}^{+0.4}$  \\
$\alpha$ & N/A & $1.89_{-0.35}^{+0.40}$  \\
$\ln 10^{10}A_s$ & $3.053_{-0.017}^{+0.016}$ & $3.057_{-0.019}^{+0.014}$  \\
$n_s$ & $0.968_{-0.003}^{+0.005}$ & $0.971_{-0.004}^{+0.003}$  \\
\hline
\end{tabular}
\caption{Constraints at 68.3\% confidence limits of the cosmological parameters for the $\alpha$-attractor and \lcdm\, models, derived using a combination of CMB, Lensing, DESI and Pantheon$+$ data.  }
\label{tab:MCMC_bf_full}
\end{table}

Finally, for the results that take into account the full data combination, we compute the Bayesian evidence (marginal likelihoods) of the $\alpha$-attractor and \lcdm\, models using the \texttt{MCEvidence} package~\cite{Heavens:2017afc}, see Table \ref{tab:MCEvidence}. This algorithm obtains the posterior for the marginal likelihood, using the $k$-th nearest-neighbor Mahalanobis distance~\cite{mahalanobis1936generalized} in the parameter space. In this analysis, we consider the $k=1$ case to minimize the effects of the inaccuracies associated with larger dimensions of the parameter space and smaller sample sizes. 

\begin{table}[!t]
\centering
\begin{tabular}{|c|c|c|c|}
\hline
Models & $\ln E_i$ & $\ln B$  & $\Delta\chi^2$ \\
\hline
\lcdm & $-2186.02$ & $-$  & $-$ \\
$\alpha$-attractor & $-2179.66$ & $6.36$ & $-0.8$ \\
\hline
\end{tabular}
\caption{The log-evidence $\ln E_i$, the log Bayes $\ln(B)$, where $B\equiv E_i/E_j$, and the differences in $\chi^2$ for the two models, using the combination of CMB, Lensing, DESI and Pantheon$+$ data.}
\label{tab:MCEvidence}
\end{table}

To evaluate the strength of evidence for or against a model in a comparison between two models using the Bayes factor $B\equiv E_i/E_j$ (the ratio of evidence between the two models), the revised Jeffreys' scale can be applied~\cite{Trotta:2008qt}. According to this scale, if $|\ln B| < 1$, the models are similar, with neither being particularly favored. When $1 < |\ln B| < 2.5$, there is weak evidence supporting one model. For $2.5 < |\ln B| < 5$, the evidence becomes moderate, and if $|\ln B| > 5$, there is strong evidence favoring one model over the other.

As can be seen in Table~\ref{tab:MCEvidence}, we find $\ln B\sim 6.36$ which implies strong evidence in favor of the $\alpha$-attractor model against \lcdm. We note that this is in agreement with the results found by DESI~\cite{DESI:2024mwx}, namely that  \lcdm~ is not preferred by the CMB+Pantheon+DESI combination. While this deviation from a constant equation of state could be attributed to new physics~\cite{Seto:2024cgo,Chudaykin:2024gol,Ye:2024ywg,Wang:2024dka,Wolf:2024stt,Li:2024qso}, it might also be due to systematics in the data~\cite{Liu:2024gfy,DESI:2024ude,DESI:2024tlb}.

An interesting point to note is that while the $\Delta \chi^2$ between the $\alpha$-attractor model and \lcdm ~is only $\Delta \chi^2\sim-0.8$, the log-Bayes is significantly larger at $\ln B\sim6.36$. To eliminate any possible issues with numerical instabilities, we further employed both a custom code and the MontePython minimizer function to estimate the minimum $\chi^2$, confirming that our results are robust. To understand the cause of having similar chi-squares but significantly different log-Bayes values, we may use the Savage-Dickey formula; for a derivation, see Appendix C of Ref.~\cite{Arjona:2024cex}. In this case, we consider $\ln B$ for the two nested models: a simpler model $M'$ with $n'$ parameters and a more general model $M$ with $n$ parameters (such that $n = n' + p$, where $p$ represents the additional parameters). In the case of such nested models, under certain mild assumptions, $\ln B$ can be analytically expressed and depends not only on the difference in chi-squares of the two models but also on the Fisher matrix of the additional parameters and their priors. In our case, the latter two factors may be what contribute to the observed difference.

We have also performed a comparison with a two parameter extension of the \lcdm~model, the so-called $w_0w_a$CDM, and summarize the results in Appendix~\ref{sec:app2}.

\section{Implication for a stochastic gravitational wave background\label{sec:GWs}}
A distinct characteristic of quintessential inflation is a post-inflation phase dominated by the kinetic energy of the scalar field. This has interesting consequences, since such a kination-dominated phase leads to Hubble evolution that differs from the standard radiation-dominated Universe, resulting in the enhancement of inflationary GWs at high frequencies.

\subsection{Inflation and post-inflation evolution}
\label{sec:post}
In order to explain the existence of DE today, the potential needs to account for the large mismatch between the inflationary plateau and the tiny energy scale of the DE. As a consequence, a crucial requirement for the scalar potential is to have a rapid fall at the end of inflation where the Universe undergoes a period of kination-dominated phase and the Hubble rate evolves as $H\propto a^{-3}$. Reheating may happen during this phase and the Universe eventually enters to radiation-dominated phase~\cite{Ford:1986sy,Chun:2009yu}, followed by the matter-dominated phase as in the standard big-bang model. Then, at later times the energy density of scalar field can dominate the Universe again, which accounts for the present DE. 

In summary, in this unified scenario of inflation and DE, the Universe undergoes periods dominated by inflaton, kination, radiation, matter, and DE. 
The duration of inflation, which is parametrized by the e-folding number $N_\ast$, can be related to the post-inflationary evolution using the following equation
\begin{eqnarray}
\label{eq:efolds2}
    N_\ast 
    &=& 67 - \ln\left(\frac{k_\ast}{a_0H_0}\right) 
    + \frac{1}{4}\ln \left(\frac{V_\ast^2}{\rho_{\rm end} M_{\rm Pl}^4}\right) \nonumber \\
    &-& \frac{1}{12}\ln g_{*,{\rm reh}} 
    + \frac{1-3w_{\rm reh}}{12(1+w_{\rm reh})}\ln \left(\frac{\rho_{\rm reh}}{\rho_{\rm end}}\right)\,,
\end{eqnarray}
where the subscripts ``$\ast$'', ``end'', ``reh'', and ``0'' indicate that the quantity is evaluated at the time when the mode $k_\ast$ exits the Hubble radius, at the end of inflation, at the transition from kination to radiation phase, and at the present time, respectively. 

For the pivot scale, we take $k_\ast=0.05\,{\rm Mpc}^{-1}$, while the third term, $V_\ast$ represents the inflaton potential at the moment when the mode $k_\ast$ exits the Hubble radius. Using Eq.~\eqref{eq:Nefold}, we find
\begin{equation}
\varphi_\ast \simeq \sqrt{\frac{3\alpha}{2}} \log\left(\frac{4 N_\ast \gamma}{3\alpha}\right)\,,
\end{equation}
which gives 
\begin{equation}
V_\ast= V(\varphi_\ast) \simeq M^2 \left(1 - \frac{3 \alpha}{2 N_\ast}\right)\,.
\label{eq:Vstar}
\end{equation}
Although the energy density of the Universe at the end of inflation, $\rho_{\rm end}$, is typically slightly lower than the inflation energy scale,  in Eq.~\eqref{eq:efolds2}, we approximate $\rho_{\rm end} \sim V_\ast$ because the correction is very small when taking the logarithm. The number of effective degrees of freedom at the reheating energy scale is assumed to be $g_{*,{\rm reh}} = 106.75$, reflecting the contribution of Standard Model particles. The last term describes the effect of the reheating epoch on the Hubble evolution; $w_{\rm reh}$ is the effective equation of state during reheating, and $\rho_{\rm reh}$ is the energy density of the Universe at the end of reheating. Assuming a kination-dominated phase during reheating, we can set $w_{\rm reh} = 1$. The energy scale at the completion of reheating, $\rho_{\rm reh}$, can be expressed in terms of the reheating temperature $T_{\rm reh}$, and the relation is given by
\begin{equation}
\rho_{\rm reh} = \frac{\pi^2}{30} \,g_{*,{\rm reh}}\,T_{\rm reh}^4\,.
\label{eq:rho_reh}
\end{equation}

The important aspect here is that now we have an implication on the value of $N_\ast$ from the CMB and LSS measurements that can be translated to implications on the post-inflationary history of the Universe through Eq.~\eqref{eq:efolds2}. In the $\alpha$-attractor model, the scalar spectral index can be directly related to the number of e-folds through Eq.~\eqref{eq:ns_r}. Thus, the 1$\sigma$ bound obtained from the analysis, $n_s=0.971^{+0.003}_{-0.004}$, indicates $N_\ast = [60.6, 76.9]$.

We also have implications on the energy scale of inflation from observation. By combining the 1$\sigma$ bound on $N_\ast$ with $\alpha=1.89^{+0.40}_{-0.35}$ and $\ln (10^{10} {\cal A}_s) = 3.057^{+0.014}_{-0.019}$, the value of $M$ is inferred to be in the range of $M^{1/2}/M_{\rm Pl} = [0.00321, 0.00411]$. Using Eq.~\eqref{eq:Vstar}, we obtain $V_\ast^{1/4}/M_{\rm Pl}=[0.00318, 0.00406]$. Finally, by combining all the constraints and assuming $w_{\rm reh} = 1$, from Eq.~\eqref{eq:efolds2}, we obtain the range of the reheating temperature as $T_{\rm reh}/{\rm GeV} = [3.5 \times 10^{-13}, 1.7 \times 10^{9}]$. 

Note that the lower bound is far beyond the other theoretical limit. While the value of $T_{\rm reh}$ depends heavily on the underlying particle physics model, and no robust prediction exists, there are two key theoretical constraints to consider. The lower bound on $T_{\rm reh}$ comes from BBN, which necessitates the completion of reheating before BBN, providing a much tighter lower bound of $T_{\rm reh} \gtrsim 10^{-3}$ GeV. The upper bound arises from the requirement that the reheating temperature must not exceed the inflation scale. Using the $1\sigma$ upper bound on $V_\ast^{1/4}/M_{\rm Pl} < 0.00403$, we find $T_{\rm reh}< 4.0\times 10^{15}$GeV. Therefore, the current bounds from CMB and LSS data do not establish a competitive lower bound; however, they do provide a relatively tight upper bound when compared to the theoretical limits. In the following subsections, we will consider the range of reheating temperature $T_{\rm reh}/{\rm GeV} =  [10^{-3}, 1.7 \times 10^9]$, where lower bound is determined by the BBN, while upper bound is derived from our CMB and LSS analysis. Then we discuss the implications of these observational bounds on the inflationary gravitational wave background. We will also demonstrate how the upper limit on high-frequency GWs, derived from BBN, can further constrain the parameter space of this model.

\subsection{GW spectrum}
\label{sec:GWspectrum}
One of the notable findings from the above discussion is that the energy scale of inflation is inferred to be in the range $V_\ast^{1/4}/M_{\rm Pl}=[0.00318, 0.00406]$, which corresponds to a relatively large tensor-to-scalar ratio $r=[0.00312, 0.00748]$. Therefore, we can expect the inflationary GW background to be detectable by next-generation CMB B-mode experiments, such as the Simons Observatory and LiteBIRD. Furthermore, the GW background could be enhanced at high frequencies due to the presence of a kination phase after inflation, making it detectable by interferometer experiments.

The amplitude of a GW background is often characterized in the form of the energy density parameter using the critical density today $\rho_{c,0} = 3M_{\rm Pl}^2 H_0^2$ as
\begin{equation}
\Omega_{\rm GW} (k) \equiv \frac{1}{\rho_{c,0}} \frac{ d \rho_{\rm GW}}{d \ln k} \,.
\end{equation}
For inflationary GWs, the modes which enter the Hubble radius during radiation-dominated phase have the spectrum of 
\begin{equation}
\Omega_{\rm GW, rad} (k)  = \frac{\Omega_{\rm rad, 0}}{12\pi^2} 
\left( \frac{g_{\ast,k}}{g_{\ast, 0}} \right)
 \left( \frac{g_{\ast s, 0}}{g_{\ast s,k}} \right)^{4/3}
 \left.\frac{H^2}{M_{\rm Pl}^2}\right|_{k=aH} \,,
 \label{eq:OGW1}
\end{equation}
where $\Omega_{\rm rad, 0}=9\times 10^{-5}$ is the energy density parameter of radiation, $H$ is the Hubble expansion rate during inflation and evaluated when the mode $k$ exits the Hubble radius. Here, the effective number of relativistic degrees of freedom $g_{*,k}$ and its counterpart for entropy $g_{*s,k}$ are evaluated when the mode $k$ enters the Hubble radius and their changes in the radiation dominated era induce step-like shapes in the GW spectrum~\cite{Watanabe:2006qe}. At high frequencies, we again use $g_{*,k}=g_{*s,k}=106.75$ assuming only Standard Model particles. The values today, labeled by $0$, are given by $g_{*,0}=3.36$ and $g_{*s,0}=3.91$.

The spectral amplitude given by Eq.~\eqref{eq:OGW1} is slightly red tilted because of the evolution of the Hubble rate during inflation and it is very often parametrized using $H_\ast^2\simeq V_\ast/(3M_{\rm Pl}^2)$ evaluated at the pivot scale $k_\ast$ and the tensor spectral tilt $n_T$. Thus, Eq.\eqref{eq:OGW1} can be written as 
\begin{equation}
\Omega_{\rm GW, rad} (k)  = \frac{\Omega_{\rm rad}}{36\pi^2}
\left( \frac{g_{\ast,k}}{g_{\ast, 0}} \right)
\left( \frac{g_{\ast s, 0}}{g_{\ast s,k}} \right)^{4/3}
\frac{V_\ast}{M_{\rm Pl}^4} \left( \frac{k}{k_\ast} \right)^{n_T}\,.
\end{equation}
The tensor tilt corresponds to the tensor-to-scalar ratio by the consistency relation as $n_T=-r/8$ and, in the case of $\alpha$-attractor model, it is given by 
\begin{equation}
n_T=-\frac{3\alpha}{2N_\ast^2} \,.
\end{equation}

If we have a kination phase before the radiation-dominated phase, GWs which entered the Hubble radius during kination phase exhibits $\Omega_{\rm GW}\propto k$ dependence and thus has a large amplitude at high frequencies. Such enhancement can be expressed by multiplying the transfer function~\cite{Duval:2024jsg}
\begin{equation}
\Omega_{\rm GW} (k)=\Omega_{\rm GW, rad} (k) ~
T(k/k_{\rm reh})^2 \,,
\end{equation}
with
\begin{equation}
T(x) =  1 - 0.5 x^{2/3} + \frac{\pi}{4} x \,,
\end{equation}
where $k_{\rm reh}$ is the wavenumber corresponding to the kination-radiation equality, given by~\cite{Figueroa:2019paj}
\begin{eqnarray}
f_{\rm reh} &=& \frac{k_{\rm reh}}{2\pi} \nonumber \\
&=& 3.8 \times 10^{-8} 
\left( \frac{g_{\ast,{\rm reh}}}{106.75} \right)^{1/2} 
\left( \frac{g_{\ast s,{\rm reh}}}{106.75} \right)^{-1/3} 
\left( \frac{T_{\rm reh}}{{\rm GeV}} \right)
{\rm Hz} \,. \nonumber \\
\end{eqnarray}
The GW spectrum has a cutoff at high frequency corresponding to the energy scale at the end of inflation $k_{\rm end}=a_{\rm end}H_{\rm end}$. If we assume $H_{\rm end}\approx H_*$, we obtain
\begin{equation}
f_{\rm max} = 2.7 \times 10^7  
\left( \frac{g_{\ast,{\rm reh}}}{106.75} \right)^{1/4}
\left( \frac{g_{\ast s,{\rm reh}}}{106.75} \right)^{-1/3} 
\left( \frac{H_\ast}{10^{13} {\rm GeV}} \right)
{\rm Hz} \,.
\label{eq:fmax}
\end{equation}

\subsection{Combined constraints}
\begin{figure}[t!]
\includegraphics[width=1\columnwidth]{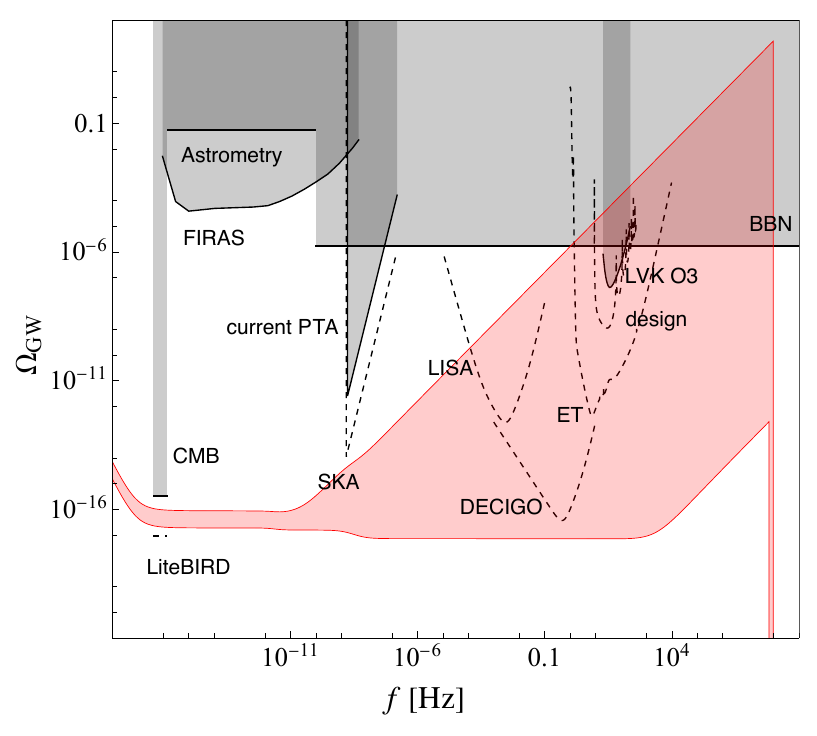}
\caption{\label{fig:GWspectrum} The spectrum of inflationary GWs predicted in the $\alpha$-attractor model with an early kination phase is compared with the sensitivities of current (solid line with gray shading) and future GW experiments (dashed line). The red shaded region indicates the possible range of the GW spectrum, with the upper and lower curves corresponding to the range of reheating temperatures, $T_{\rm reh}/{\rm GeV} = [10^{-3}, 1.7 \times 10^9]$. These bounds reflect the theoretical lower limit necessary for successful BBN and the observational upper limit (1$\sigma$) indicated by CMB and LSS data. }
\end{figure}

\begin{figure}[t!]
\includegraphics[width=1\columnwidth]{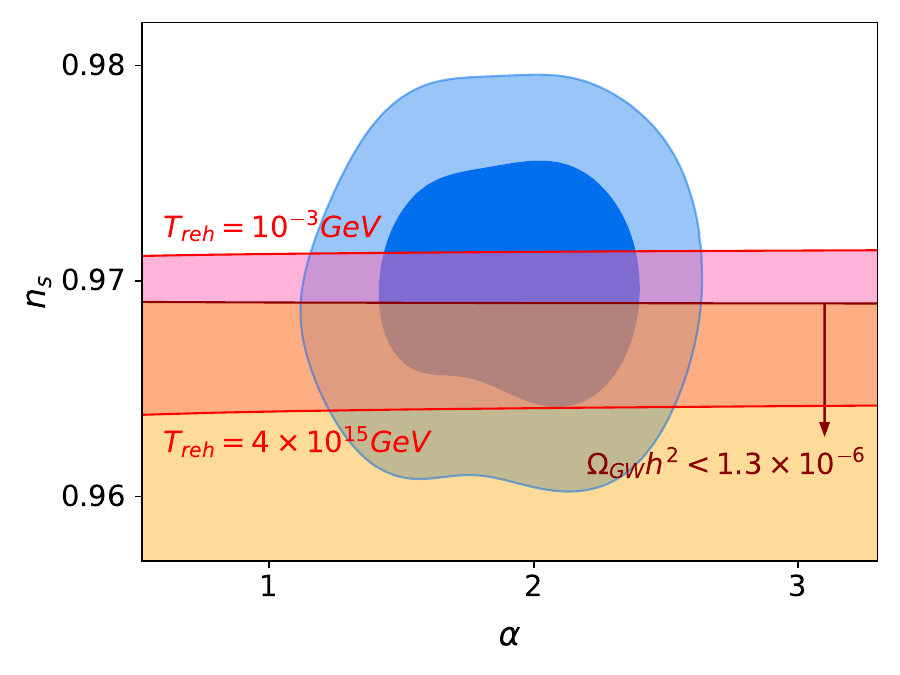}
\caption{\label{fig:GWconstraint} The constraint in the $\alpha$--$n_s$ plane, obtained by CMB+Lensing, DESI and Pantheon+ likelihoods (blue contour, same as in Fig.~\ref{fig:full}), is plotted alongside the bounds from the GW amplitude at high frequencies and the theoretical limits on reheating temperature.  The orange shaded area indicates the parameter region constrained by the BBN bound, $\Omega_{\rm GW} h^2 < 1.3 \times 10^{-6}$. The red band indicates the region predicting a reasonable value of the reheating temperature, $T_{\rm reh}/{\rm GeV} = [10^{-3}, 4.0 \times 10^{15}]$.}
\end{figure}

By using the indicated range obtained from the MCMC analysis, $V_\ast^{1/4}/M_{\rm Pl}=[0.00318, 0.00406]$ and $T_{\rm reh}/{\rm GeV} = [10^{-3}, 1.7 \times 10^9]$, discussed in Sec.~\ref{sec:post}, in Fig.~\ref{fig:GWspectrum}, we show the GW spectrum using the formulas described in Sec.~\ref{sec:GWspectrum}. 
\SK{\footnote{It is worth noting that there is also flexibility to adjust the parameters $w_{\rm reh}$ and $g_{*,{\rm reh}}$, both of which influences the GW spectral amplitude. The values adopted here $w_{\rm reh}=1$ (the value of kination domination) and $g_{*,{\rm reh}}=106.75$ (the value from the standard model) are well-motivated, but in principle, they could be set as free parameters.}}
Notably, when the reheating temperature is low $T_{\rm reh} \lesssim 10^5$GeV (corresponding to large e-folding number, $N_\ast \gtrsim 64$), the model predicts too large GW amplitude at high frequencies, which violates constraints from the BBN and LIGO-Virgo-KAGRA (LVK) O3 upper bound~\cite{KAGRA:2021kbb}, shown by shaded gray region. 
Here, the BBN bound is based on the requirement that the energy density of GWs should not be too large, as it would alter the Hubble expansion rate and disrupt the successful BBN. The recent joint CMB+BBN analysis implies $\Omega_{\rm GW} h^2 < 1.3 \times 10^{-6} $ at $2 \sigma$ for $f > 2 \times 10^{-11}$Hz~\cite{Yeh:2022heq}.

This highlights the interesting fact that certain parameter regions can be further constrained by GW constraints. In Fig.~\ref{fig:GWconstraint}, the parameter space allowed by the BBN bound $\Omega_{\rm GW} h^2 < 1.3 \times 10^{-6}$ is shaded with orange. As indicated in Fig~\ref{fig:GWspectrum}, the parameter space predicting too low reheating temperature (larger e-folding number, larger $n_s$) is excluded as it predicts too large GW spectrum at high frequencies. Furthermore, the red shaded region indicates where the parameters predict a reasonable reheating temperature. A too low reheating temperature ($T_{\rm reh} \lesssim 10^{-3}$GeV) disrupts the successful BBN, while an excessively high reheating temperature ($T_{\rm reh} \sim 4.0\times 10^{15}$GeV) implies instant reheating\footnote{Note that in the context of supergravity, considerations on the overproduction of gravitinos can further constrain the reheating temperature to $T_{\rm reh} \lesssim  10^8 - 10^9$GeV~\cite{Kawasaki:2006gs,Dimopoulos:2017zvq} or even down to $10^6$GeV~\cite{Kohri:2005wn}.}. This temperature corresponds to the e-folding number of $N_\ast\sim 56$ and when the e-folding is lower than it we need a phase with $w<1/3$ such as a early matter domination, which requires extra physics to achieve it in the picture of $\alpha$-attractor model. Therefore, the overlap of the red and orange shaded areas represents the allowed region, enabling us to reject more parameter space of the model. We also find that the CMB and LSS constraints indicate a slight preference for lower reheating temperatures; however, higher reheating temperatures are still allowed at the $2\sigma$ level.

However, we note that this is a simplified analysis, as it assumes an instant transition from inflation to the kination phase. In reality, a smooth transition is expected, which would smooth out the peak feature at the highest frequency, a critical aspect when discussing the BBN bound. In addition, higher-order terms in the slow-roll approximation (such as the tensor runnings) become important at such high frequencies~\cite{Kuroyanagi:2008ye,Kuroyanagi:2011iw}. Furthermore, the prediction of the e-folding number is influenced by such smooth transitions. For a more precise prediction, a detailed numerical analysis simulating the dynamics of the scalar field from inflation to the present day would be required.

Another point to note for Fig~\ref{fig:GWconstraint} is that, to plot the curves for the GW amplitude and reheating temperatures, we fixed the other parameter values to their marginalized mean values as shown in Table \ref{tab:MCMC_bf_full}, while the blue contours were obtained by marginalizing over the errors in other parameters. The other parameters, particularly $A_s$, cause slight modifications in the GW amplitude and reheating temperature. Therefore, a proper likelihood analysis is necessary for a fair comparison with the CMB and LSS data. Nonetheless, we have confirmed that changes in these parameters at the $2-3\sigma$ level do not induce significant variations in the figure.


\section{Conclusions \label{sec:conclusions}}
In this work, we explored the compatibility of a specific class of $\alpha$-attractor models when using the latest observational data, motivated by recent implications of evolving DE equation of state parameter from DESI.
The $\alpha$-attractor models present a compelling and unified framework, motivated by high-energy physics and supergravity, as an alternative to traditional single-field slow-roll inflation. These models connect early- and late-time physics, offering a viable and physically motivated alternative to other phenomenological models.

Specifically, in our analysis we considered the latest cosmological data, including the Planck 2018 CMB temperature and polarization spectra, the CMB lensing, the latest DESI BAO measurements, the Pantheon+ SnIa data. After performing a standard MCMC analysis, we presented our main results in Tables~\ref{tab:MCMC_bf_cmb} and \ref{tab:MCMC_bf_full} and the $68.3 \%$--$95.5 \%$ confidence contours for both models in  Figs.~\ref{fig:cmb} and \ref{fig:full}. 

We found that the constraints on the  parameters of both \lcdm\, and the $\alpha$-attractor models are comparable, while when all data are considered the $\alpha$ parameter itself is constrained to be $\alpha\simeq 1.89_{-0.35}^{+0.40}$. Out of the values of $3\,\alpha \in \{1,2,3,4,5,6,7\}$ that are theoretically motivated by supergravity, we find that the data seem to prefer (rounding to the closest integer) either $\alpha=5/3$ or $\alpha=2$ within the $68.3 \%$ confidence limit.

As mentioned, the reason for this is that the particular $\alpha$-attractor model we considered roughly mimics, within the allowed parameter space for quintessence $w(z)\ge-1$ as it cannot cross the phantom divide line (see for example Ref.~\cite{Nesseris:2006er}), an equation of state with $(w_0,w_a)$ values that are in agreement with the ones found by DESI, as seen in our Fig.~\ref{fig:attractorw0wa} and the right panel of Fig. 6 of Ref.~\cite{DESI:2024mwx}. 

We also considered a connection of the $\alpha$-attractor models with GWs by comparing the spectrum of inflationary GWs predicted in the $\alpha$-attractor model with early kination phase; see Fig.~\ref{fig:GWspectrum}, compared with sensitivities of current/future GW experiments. 
We find the interesting fact that certain parameter regions can be further constrained by the current BBN and CMB constraints on the GW amplitude, see Fig.~\ref{fig:GWconstraint}. These constraints could be strengthened by future CMB B-mode measurements as well as constraints by interferometer experiments. Additionally, by considering an appropriate range of reheating temperatures, we demonstrated that the parameter space can be further tightened. 
However, the analysis we performed here is only indicative, thus we leave a more complete study for future work.

Overall, we find that the $\alpha$-attractor models seem to provide an attractive alternative to the cosmological constant model, providing a deep connection between early- and late-time physics, along with a plethora of observables that can be tightly constrained either via LSS or GW data. Thus, future observations, especially using GWs as noted in our work, will be instrumental in further testing the model.


\section*{Acknowledgements}
The authors thank Nils Sch{\"o}neberg for useful discussions and they acknowledge the use of the Finis Terrae III supercomputer, which is part of the Centro de Supercomputacion de Galicia (CESGA) and is funded by the Ministry of Science and Innovation, Xunta de Galicia and ERDF (European Regional Development Fund). The authors also acknowledge support from the research project PID2021-123012NB-C43 and the Spanish Research Agency (Agencia Estatal de Investigaci\'on) through the Grant IFT Centro de Excelencia Severo Ochoa No CEX2020-001007-S, funded by MCIN/AEI/10.13039/501100011033. GA's and SK's research is supported by the Spanish Attraccion de Talento contract no. 2019-T1/TIC-13177 and 2023-5A/TIC-28945 granted by the Comunidad de Madrid. SK is partly supported by the I+D grant PID2020-118159GA-C42 and PID2023-149018NB-C42 funded by MCIN/AEI/10.13039/501100011033, the Consolidaci\'on Investigadora 2022 grant CNS2022-135211, and Japan Society for the Promotion of Science (JSPS) KAKENHI Grant no. 20H01899, 20H05853, and 23H00110. MC acknowledges support from the ``Ram\'on Areces'' Foundation through the ``Programa de Ayudas Fundaci\'on Ram\'on Areces para la realizaci\'on de Tesis Doctorales en Ciencias de la Vida y de la Materia 2023''.

\begin{figure*}[t!]
\includegraphics[width=1\columnwidth]{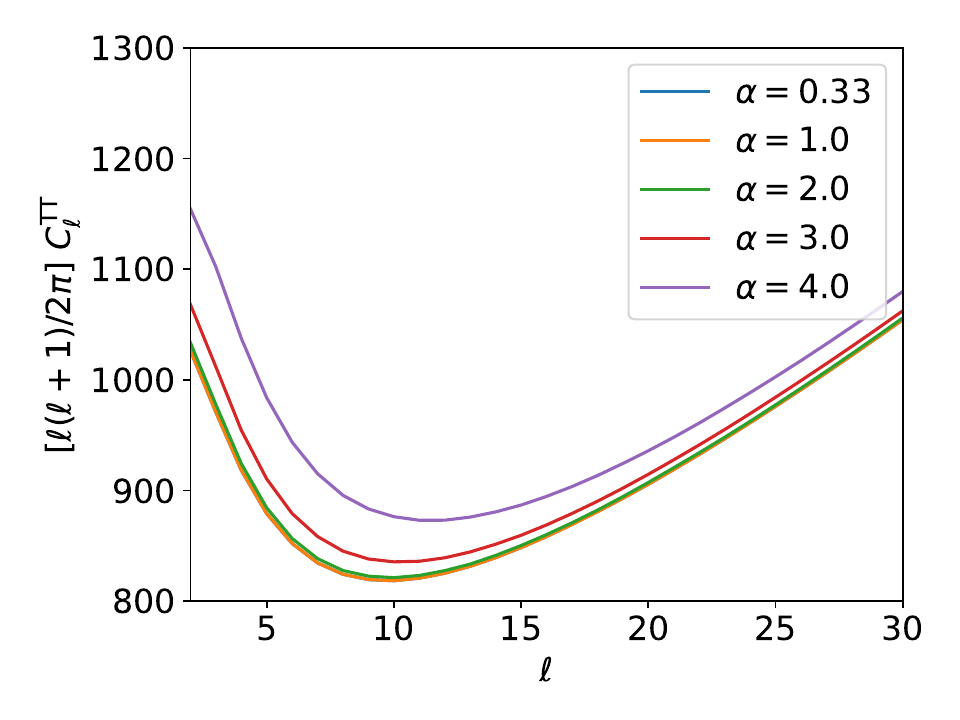}
\includegraphics[width=1\columnwidth]{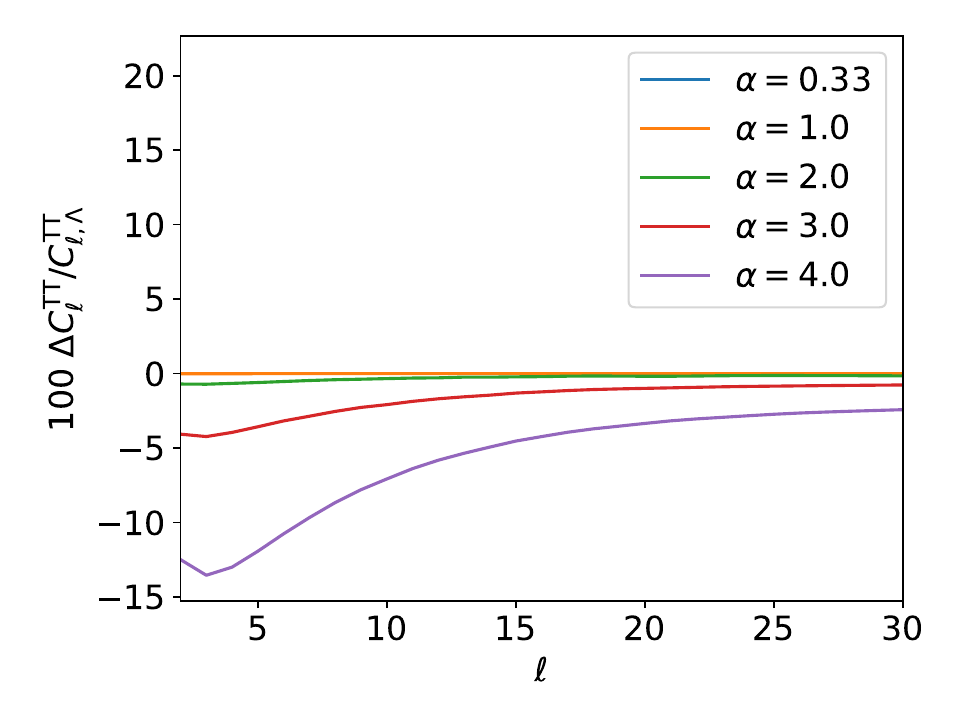}
\caption{\label{fig:ISW} The CMB TT spectra for the $\alpha$-attractor model for $\alpha = \{1/3,1,2,3,4\}$ at $\ell <30$ (left) and the percent difference (right). The maximum deviation is $\sim12\%$ for $\alpha=4$.}
\end{figure*}

\begin{figure*}[t!]
\includegraphics[width=1\columnwidth]{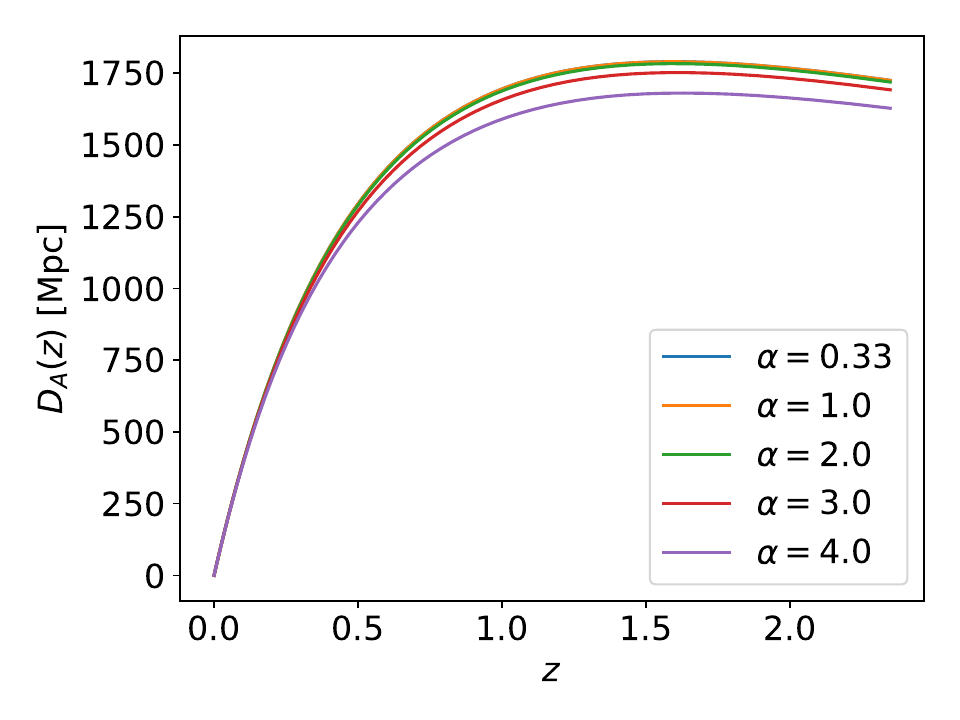}
\includegraphics[width=1\columnwidth]{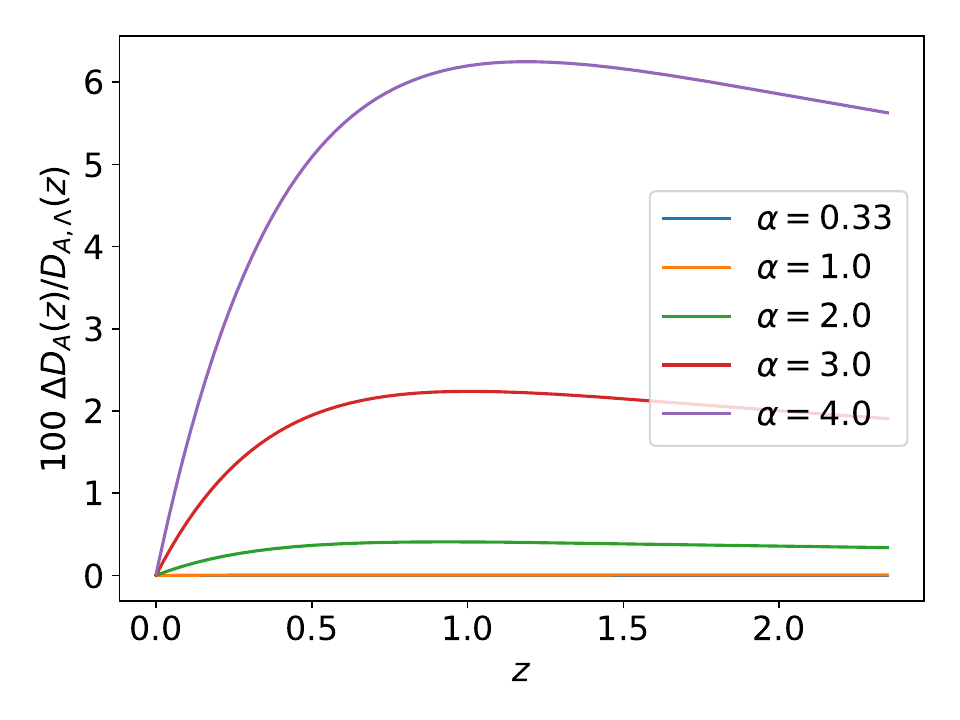}
\caption{\label{fig:DA} The angular diameter distance for the $\alpha$-attractor model for $\alpha = \{1/3,1,2,3,4\}$ at $z <2.5$ (left) and the percent difference (right). The maximum deviation is $\sim6\%$ for $\alpha=4$.}
\end{figure*}

\begin{figure*}[t!]
\includegraphics[width=1\columnwidth]{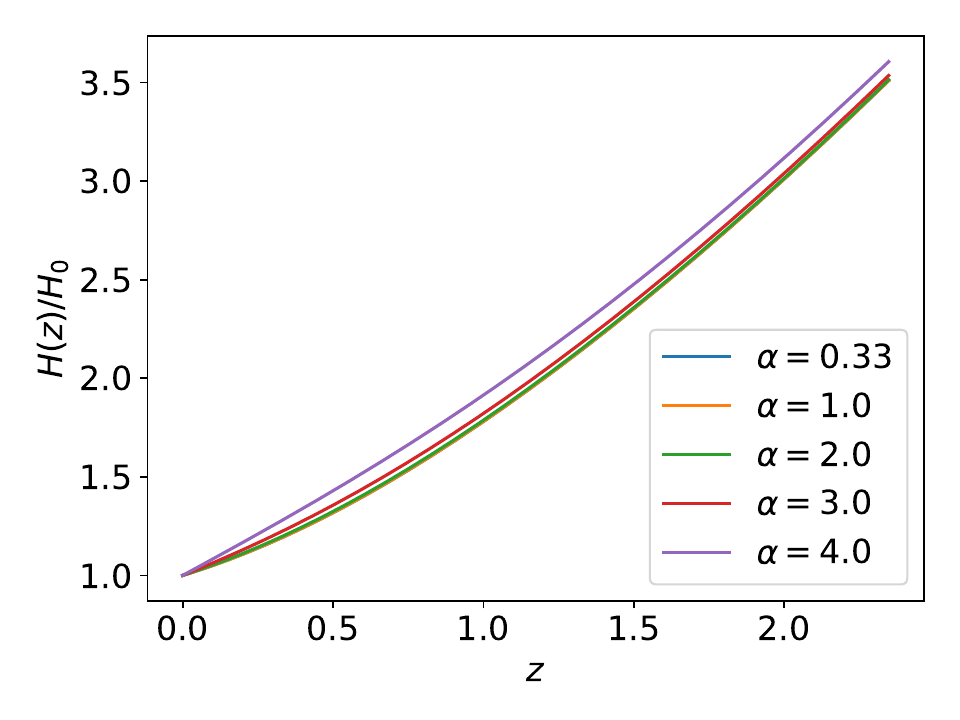}
\includegraphics[width=1\columnwidth]{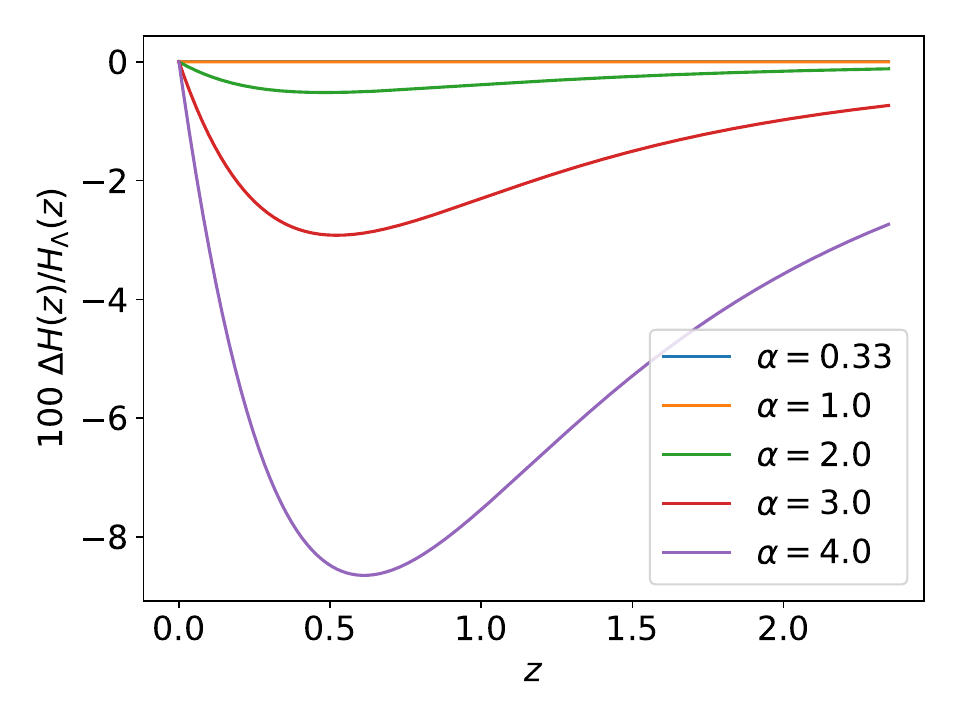}
\caption{\label{fig:H} The normalized Hubble parameter for the $\alpha$-attractor model for $\alpha = \{1/3,1,2,3,4\}$ at $z <2.5$ (left) and the percent difference (right). The maximum deviation is $\sim8\%$ for $\alpha=4$.}
\end{figure*}

\begin{figure}[t!]
\vspace{-0.5cm}
\includegraphics[width=1\columnwidth]{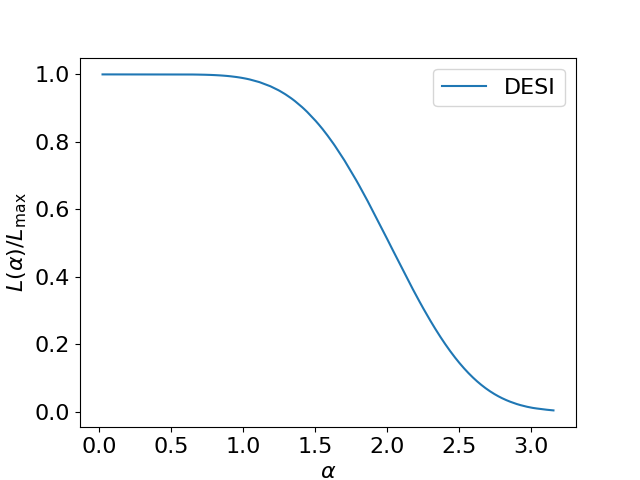}
\caption{\label{fig:desi} The 1D marginalized likelihood for the $\alpha$ parameter, using the DESI likelihood alone.}
\end{figure}

\appendix
\section{Quantitative discussion of the $\alpha$-attractor models \label{sec:app}}
The main effect of $\alpha$-attractor models at late times, compared to \lcdm, is a change at the background expansion of the Universe causing deviations at a) the CMB spectra at low multipoles via the Integrated Sachs-Wolfe (ISW) effect, and b) at the angular diameter distance, affecting directly the BAO measurements. 

In Fig.~\ref{fig:ISW}, we show the CMB TT spectra for values of $\alpha \in [1/3,4]$ assuming all other parameters are fixed (left) and the percent deviation from the $\alpha=1/3$ case (right panel), which is practically indistinguishable from \lcdm. As can be seen, the largest deviation for the TT spectra is for $\alpha=4$ and reaches $\sim 12 \%$ at low $\ell$. 

In Fig.~\ref{fig:DA}, we show the angular diameter distance $D_A$ as a function of the redshift $z$ (left) for the same parameters as before and the percent deviation from the $\alpha=1/3$ case (right panel). In this case, the maximum deviation reaches $\sim 6\%$ for $\alpha=4$ at $z\sim1.2$. The luminosity distance is expected to show a similar behavior due to the duality relation, which we assume it holds in these models. Similarly in Fig.~\ref{fig:H} we show the normalized Hubble parameter $H(z)/H_0$ as a function of the redshift $z$ (left) and the percent deviation (right). Here, the maximum deviation is $\sim8\%$ for $\alpha=4$ at $z\sim0.5$. Thus, overall we expect the BAO constraints to be (in principle) stronger at intermediate redshifts, approximately at $z\in[0.5,1.5]$. For completeness, in Fig.~\ref{fig:desi}, we also show the marginalized 1D constraints on $\alpha$, using the DESI data only. As can be seen, in this case we only get a bound of $\alpha \lesssim 3$. 

\section{Comparison with the $w_0w_a$CDM model \label{sec:app2}}
\begin{figure*}[t!]
\includegraphics[width=1\textwidth]{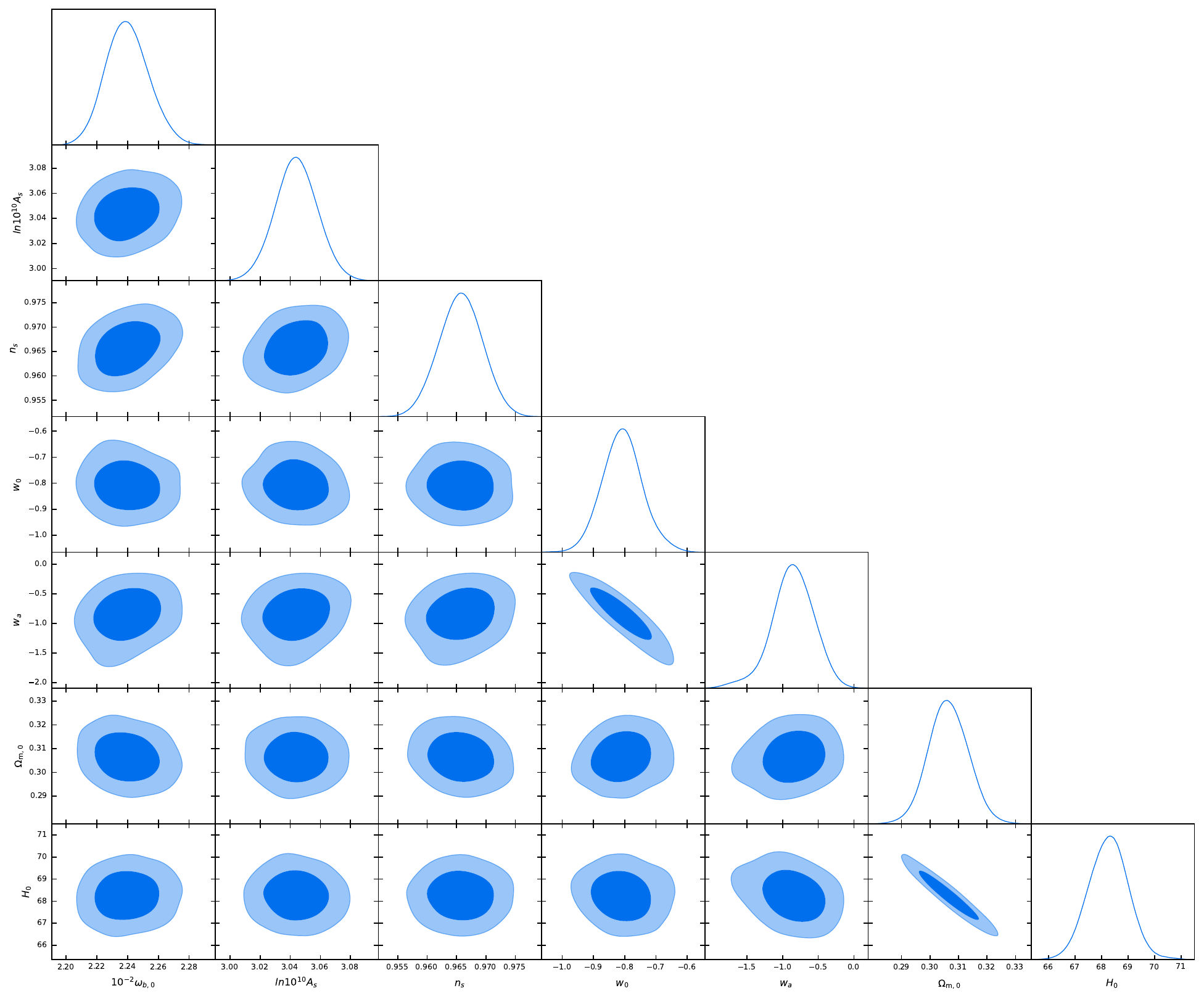}
\caption{\label{fig:cpl}The $68.3 \%$--$95.5 \%$ confidence contours for the cosmological parameters of the CPL model, which is derived using CMB$+$Lensing, SnIa and DESI data.}
\end{figure*}

In this section we also briefly present some constraints on a two-parameter extension of the \lcdm~model, in terms of the dark energy equation of state parameter, of the form $w(a)=w_0+w_a\,(1-a)$, where $w_0$ and $w_a$ are to free parameters. As before, using the CMB+Lensing, SnIa and DESI data we find the constraints as shown in Fig.~\ref{fig:cpl}, and compare the results with the ones obtained for the $\alpha$-attractor model.

We observe that the best fit parameters of the $w_0w_a$CDM model, are $w_0=-0.796_{-0.067}^{+0.061}$ and $w_a=-0.972_{-0.270}^{+0.300}$ respectively, which are in very good agreement with the ones reported in Ref.~\cite{DESI:2024mwx}. The log-evidence for this model is $\ln E_i = -2172.89$ and the log Bayes compared to \lcdm~(see Tab.~\ref{tab:MCEvidence}) is $\ln B = 13.13$, which shows strong overall preference for the $w_0w_a$CDM model. 

\bibliographystyle{apsrev4-1}
\bibliography{Bibliography}

\end{document}